\documentstyle[12pt]{article}
\begin{document}
\centerline{\bf {TIME DEPENDENCE OF BRANS-DICKE PARAMETER  $\omega$}} 
\centerline{\bf { FOR AN EXPANDING UNIVERSE}}

\vspace{0.2in}
\centerline{  B.K.Sahoo$^{1}$ and L.P.Singh$^{2}$}
\centerline{ Department of Physics,} 
\centerline{Utkal University,Bhubaneswar-751004,India}
\centerline{$^{1}$    bijaya@iopb.res.in}
\centerline{$^{2}$   lambodar@iopb.res.in}
\vspace{0.25in}
\begin{abstract}
  We have studied the  time dependence of  $\omega$ for an expanding universe in the generalised B-D theory and have obtained its explicit dependence on the  nature of matter contained in the universe  in different era.Lastly,  we discuss how the observed accelerated expansion  of the present universe can be accomodated in the formalism.
\end{abstract}
\vspace{0.25in}

\vspace{0.25in} 
PACS NO:      98.80.-k,98.80.cq
\newpage
\section{INTRODUCTION}

 The Brans-Dicke theory is defined by a constant coupling function
 $\omega$  and a scalar field  $\phi$  .The relative importance is 
determined by the arbitary coupling function $ \omega $ [1].The
generalised B-D theory $ [2]$ involves an extension  of original theory to the case of a time dependent  coupling function  $ \omega = \omega (t)$. Generalised B-D model  is  special for more than one reason .
It appears naturally in super gravity theory, Kaluza-Klein theories
and in all the known effective string actions .It is perhaps the most
natural  extension of General Relativity $[3]$,which may explain its
obiquitous appearance in fundamental theories.In the generalised B-D
theory,sometimes referred  to as graviton-dilaton theory, $\omega$  is an arbitary function of the  scalar field $\phi$  (dilaton) .Hence it includes a  number of models, one for every function   $\omega $  .GR is obtained when the field  $\phi$   = constant and $ \omega (t)=\infty$ $ [4,5]$.
                 
              Brans-Dicke theory with a constant $\omega$ is a successful   theory because it explains  almost all the important
  features of the evolution of the Universe. Some of the problems like
  inflation $[6]$, early and late time behaviour of Universe
  $[7]$, cosmic acceleration  and structure formation $[8]$, cosmic
  acceleration,quintessence and coincidence problem $[9]$, self
  -interacting potential and cosmic acceleration $[10]$  can be
  explained in the  B-D formalism .  For large  $\omega$   B-D  theory
  gives the correct amount of inflation and early and  late time
  behaviour, and for small negative  $\omega$  it correctly explains
  cosmic  acceleration, structure formation and coincidence problem .
      
          Brans-Dicke theory with a time dependent $\omega$ is also
          an interesting theory in its own right . Not only it gets a strong
  support from string and Kaluza-Klein theories, a few attempts have also been  taken using this formalism to study the dynamics of the
  Universe . However, all these attempts address the problems of
  evolution of universe $[11]$, cosmic accleration  and quintessence $[9]$ etc. in a
  qualitative  way without exibiting the explicit time dependence of
  $\omega$ . Also,Alimi and  Serna $[5]$ have shown that this  time varing
  $\omega$  theory includes a number of models; one each for  every
  time dependence of $\omega$ . So it is of natural interest to seek
  exact derivation of time  dependence of $ \omega $ from the
  dynamical equations of the Universe .
   
              Our aim, in this paper therefore, is to derive explicit
              time dependence of $ \omega $ for simple expanding
              solutions of field and  wave equations in the
              Brans-Dicke theory for all era which can give some
              information regarding the early and late time behaviour
              of  the universe .This will also provide  the oppertunity 
              to examine if the derived time-dependence of $\omega$ can be              used to  explain atleast some of
              the presently observed properties of the universe like
              cosmic accleration, structure formation etc.

\section{ THE GENERAL $ \omega $ }
             For a  Universe filled with perfect fluid and described by Friedmann-Robertson -Walker space-time with scale
             factor $a(t)$ and  spatial curvature index
 $k$, the gravitational
 field equations in B-D theory
 with time dependent $\omega$, are 
\begin{equation}
\frac{\dot{a}^{2}+k}{a^{2}}+\frac{\dot{a}\dot{\phi}}{a\phi}-
\frac{\omega \dot{\phi}^{2}}{6\phi^{2}} =\frac{\rho}{3\phi}  
\end{equation}

\begin{equation}
2\frac{\ddot {a}}{a}+\frac{\dot{a}^{2}+k}{a^{2}}+
 \frac{\omega{\dot{\phi}^{2}}}{2\phi^{2}}+
 2\frac{{\dot{a}}{\dot{\phi}}}{a\phi}+\frac{{\ddot{\phi}}}{\phi}
 =-\frac{P}{\phi}
\end{equation}
where $\rho$ and P are  respectively the energy density and
 pressure of the fluid distribution . The equation of state of
 the fluid is given by$P=\gamma\rho$. Some of the values of $\gamma$ 
  for typical cases are -1 (vacuum), 0 (dust), 1/3 (radiation),
1(massless scalar field) .

           The wave equation for Brans-Dicke scalar field when $\omega$  is a function of time is $ [2,3,4,5,]$

\begin{equation}
 \ddot{\phi}+3\frac{\dot{a}\dot{\phi}}{a}=\frac{\rho-3P}{2\omega+3}-\frac{\dot{\omega}\dot{\phi}}{2\omega+3}.
\end{equation}

 Energy conservation equation which can be obtaioned from eqs.$(1)$,$(2)$,
and$(3)$ is,

\begin{equation}
\dot{\rho}+3\frac{\dot{a}}{a}\left(\rho+P\right)=0.
\end{equation}

 One important property of Brans-Dicke theory is that it gives simple expanding solutions for field $\phi(t)$ and scale factor $a(t)$ which are compatible with solar system experiments $[12]$. To derive the  time dependence of  $\omega$ which satisfies both field and wave equations, we assume the time dependence of the scale factor and scalar field in the following form which provides simple  expanding solutions,
 
\begin{equation}
a=a_{0}\left(\frac{t}{t_{0}}\right)^{\alpha}.
\end{equation}

\begin{equation}
\phi=\phi_{0}\left(\frac{t}{t_{0}}\right)^{\beta}.
\end{equation}

\noindent $a_{0}$ and $\phi_{0}$ are the present values of $a(t)$and $\phi(t)$.
 From eq.(1) and taking $k=0$ to be consistent with the paradigm of
 inflation, we get,\\

\centerline{$\left(\frac{\dot{a}}{a}\right)^{2}+
\frac{\dot{a}\dot{\phi}}{a\phi}-\left(\frac{\rho}{3\phi}+
\frac{\omega(t)\dot{\phi}^{2}}{6\phi^{2}}\right)=0$}.

\noindent This immediatly leads to\\
\centerline{ $\frac{{\dot{a}}}{{a}}$=$-\frac{\dot{\phi}}{2\phi}\pm
\left[\frac{(2\omega+3)\dot{\phi}^{2}}{12\phi^{2}}+
\frac{\rho}{3\phi}\right]^{1/2}$}.\\

\noindent This equation can alternatively,be written in the form \\
\begin{equation}
\left[\left(\frac{\dot{a}}{a}+\frac{\dot{\phi}}{2\phi}\right)^{2}-
\frac{\Omega(t)\dot{\phi}^{2}}{12\phi^{2}}\right]3\phi=\rho,
\end{equation}

\noindent where $\Omega (t)=2\omega(t) +3$ .\\
Solving eq.$(4)$ one gets
\begin{equation}
\rho=\rho_{0}a^{-3\left(1+\gamma\right)}.
\end{equation}

\noindent Using eq.$(5),(6)$ and $(8)$ in  $(7)$ we get \\

$\left[\left(\frac{2\alpha+\beta}{2t}\right)^{2}-\frac{{\Omega(t)\beta^{2}}}{{12t^{2}}}\right]3\phi_{0}\left(\frac{t}{t_{0}}\right)^{\beta} = \rho_{0}a_{0}^{-3\left(1+\gamma\right)} \left(\frac{t}{t_{0}}\right)^{-3\alpha\left(1+\gamma\right)}$.\\

\noindent From the above equation, one immediatly obtains,\\

\begin{equation}
\Omega(t) =- \frac{4\rho_{0}t_{0}^{2}a_{0}^{-3\left(1+\gamma\right)}}{\beta^{2}\phi_{0}} \left(\frac{t}{t_{0}}\right)^{-3\alpha(1+\gamma)-\beta+2} + 3\left(\frac{2\alpha+\beta}{\beta}\right)^{2}. 
\end{equation}

\noindent Since $\Omega (t)$ =2$\omega (t)$ +3 ,we are led to the time dependence of $\omega$ of the form,\\

\begin{equation}
\omega(t) = - \frac{2}{\beta^{2}} \frac{\rho_{0}t_{0}^{2}a_{0}^{-3(1+\gamma)}}{\phi_{0}} \left(\frac{t}{t_{0}}\right)^{-3\alpha(1+\gamma)-\beta+2}.
\end{equation}

We now proceed to check the consistency of above time- dependence of
  $\omega$ with eq.$(3)$ which is the wave equation for the scalar field
$\phi(t)$.
Using eqs. $(5),(6),(8)$ and $(9)$,  eq.$(3)$ reduces to\\

$\left[\beta\left(\beta-1\right)+3\alpha\beta\right]t^{\beta-2} =
\frac{\left[\left(1-3\gamma\right)-\frac{2}{\beta}\left(3\alpha\left(1+\gamma\right)+\beta-2\right)\right]t^{-3\alpha\left(1+\gamma\right)}}{-\frac{4}{\beta^{2}}t^{-3\alpha\left(1+\gamma\right)-\beta+2}}$.\\

   We  neglect  $3\left(\frac{2\alpha+\beta}{\beta}\right)^{2}$ in the denominator as we are interested in time dependence only.
Thus above equation reduces to\\
\begin{equation}
\beta\left[\frac{3\left(1-\gamma\right)}{4}\beta+\frac{3\alpha\left(1-\gamma\right)}{2}\right]=0.
\end{equation}

Eq.(11) has obvious solution 
$\beta=0$ or $\beta=-2\alpha$.
Thus, eq.(9) will satisfy both field and wave equations when $\beta=0$ or,
$\beta=-2\alpha$.It may be noted in passing that for these two cases in particular the neglect of the term  $3(\frac{2\alpha+\beta}{\beta})^{2}$ as noted earlier is evidently justified. We now consider these two cases one by one.\\

\underline{CASE -  1}:    $\beta=0$.\\

         For $\beta=0$  eq.(10) gives $\omega =\infty$, eq.(7) gives $\phi=\phi_{0}=$ constant and $a=a_{0}\left(\frac{t}{t_{0}}\right)^{\alpha}$.
Thus, for $\beta=0$ Brans- Dicke model goes over to General Relativity 
 $[5]$.Here,$\alpha$ is not related to $ \beta$ and its value can be obtained by solving 
equations of General relativity $[13]$.\\

\underline{CASE - 2}:      $\beta=-2\alpha$.\\

Eq.$(10)$ in this case reduces to,

\begin{equation}
\omega(t) = - \frac{1}{2\alpha^{2}}  \frac{\rho_{0} t_{0}^{2} a_{0}^{-3(1+\gamma)}}{\phi_{0}} \left(\frac{t}{t_{0}}\right)^{-3\alpha(1+\gamma)+2\alpha+2}.
\end{equation}

 For completeness we also write,\\
 \begin{equation}
a=a_{0}\left(\frac{t}{t_{0}}\right)^{\alpha}.
\end{equation}

\begin{equation}
\phi=\phi_{0}\left(\frac{t}{t_{0}}\right)^{-2\alpha}. 
\end{equation}

   Thus, to work within the B-D model with a time-dependent $\omega$ we have to  take $\beta=-2\alpha$  leading to the above solutions. Clearly,  for a given  $\alpha >0$ (non-negative and non-zero )which can be obtained from the observational data $[12]$ the time dependence of   $a(t)$, $\phi(t)$ and $\omega(t)$ are fixed through the above equations.

\section{$ \omega $ FOR DIFFERENT  ERA}

 It is clear from eq.$(12)$ that, the parameter $\gamma$ which takes
 different values in  different  era,  controls the time  dependece of  $\omega$ in 
the respective era as detailed below.

\subsection{Vacuum dominated era:  $ \gamma=-1$}
 Here,

\begin{equation}                     
\omega = - \frac{1}{2\alpha^{2}} \frac{\rho_{0}t_{0}^{2}}{\phi_{0}} \left(\frac{t}{t_{0}}\right)^{2\alpha+2}.
\end{equation}

 Since for an expanding universe $\alpha>0$, the time dependence of $\omega$ will always be governed by power of time greater than $2$. Hence,
 $\omega$  decreases faster than $ -t^{2}$ with time.

\subsection{ Radiation dominated era:    $\gamma=1/3$}.\\
  Here, 
\begin{equation}
\omega=-\frac{1}{2\alpha^{2}}\frac{\rho_{0}t_{0}^{2}}{a_{0}^{4}
\phi_{0}}\left(\frac{t}{t_{0}}\right)^{-2\alpha+2}.
\end{equation}.

  Here, again $\omega$ is a  decreasing function of time as the
  Universe undergoes a deccelerated expansion in this era with  $0 <\alpha <1 $.

\subsection{Matter dominated era :$\gamma=0$}
Here,

\begin{equation}
 \omega = - \frac{1}{2\alpha^{2}}\frac{\rho_{0}t_{0}^{2}}{a_{0}^{3}\phi_{0}}
\left(\frac{t}{t_{0}}\right)^{-\alpha+2}.
\end{equation}

 The implications of this time dependence is dicussed in the  next  section.

\subsection{Massless scalar field dominated era :$\gamma =1$}
   Here,
\begin{equation}
\omega = - \frac{1}{2\alpha^{2}} \frac{\rho_{0}t_{0}^{2}}{a_{0}^{6}\phi_{0}} \left(\frac{t}{t_{0}}\right)^{-4\alpha+2}.
\end{equation}

\section{$ \omega$ FOR PRESENT  UNIVERSE.}

 The present observable universe contains cold matter of negligible  pressure 
(dust). We,therefore, take the time dependence of  $\omega$  as given by eq. $(17)$i.e,

\begin{equation}       
 \omega=-\frac{1}{2\alpha^{2}}\frac{\rho_{0}t_{0}^{2}}{a_{0}^{3}\phi_{0}}\left(\frac{t}{t_{0}}\right)^{-\alpha+2}.
\end{equation}

\noindent For present time,taking $ t=t_{o}$, one gets,
 
\begin{equation}  
\omega=-\frac{1}{2\alpha^{2}}\frac{\rho_{0}t_{0}^{2}}{a_{0}^{3}\phi_{0}}=
\omega_{0},
\end{equation}

\noindent where $\omega_{o}$  represents present value of $\omega$.
    $\rho_{0}$,$\phi_{0}$,$a_{0}$ are positive non-zero constants.  However,
 they can all be set equal to  $1$  with no loss of  generality if time  t is measured in units of  $\sqrt{\frac{\phi_{0}a_{0}^{3}}{\rho_{0}}}$ .Therefore,setting
 $\frac{\rho_{0}t_{0}^{2}}{a_{0}^{3}\phi_{0}} =1$,we get, 

\begin{equation}
 \omega _{0}=-\frac{1}{2\alpha^{2}}.
\end{equation}

For the presently observed acceleration to be accomodated $\alpha$
needs to be greater than 1.In that case  $\omega_{0}$ as given by
eq.$(21)$ has the minimum value of $-\frac{1}{2}$ .This result is in
agreement with the observation made by  Banerjee and Pavon $[9]$  that
$\omega$ value must be greater than $-\frac{3}{2}$ for  Newtonian
constant of gravitation G and the scalar field energy density to
remain positive .\\

             We now recall that the present observational data for decceleration parameter$ [12]$ is\\
\begin{equation}
               -1 < q_{0} < 0.
\end{equation}

\noindent Since by defination,

\begin{equation}
               q_{0}=-\frac{\ddot{a}a}{\dot{a}^{2}}.
\end{equation}
 \noindent Using eq.$(5)$ we get,

\begin{equation}
                   q_{0} =-\frac{\alpha-1}{\alpha}.
\end{equation}
 
 \noindent We see from eq. $(22)$  and eq. $(24)$ that in order to obtain $q_{0}$ 
between   -1   and    0 ,      $\alpha$  need only be greater than $1$.
  Putting $\alpha =1+\epsilon$, where   $\epsilon > 0$ but small, the present
  day solutions of a $\omega$-varying B-D theory can be
  taken as\\ 
\begin{eqnarray}
a=a_{0}\left(\frac{t}{t_{0}}\right)^{1+\epsilon}.\\
\phi=\phi_{0}\left(\frac{t}{t_{0}}\right)^{-2\left(1+\epsilon\right)}.\\
\omega = - \frac{1}{2(1+\epsilon)^{2}} \left(\frac{t}{t_{0}}\right)^{1-\epsilon}.
\end{eqnarray}

    Thus, we find that time dependence of $\omega$ obtained through
    consistent solutions 
of B-D field equations and the wave equation leads to a negative  constant $
\omega$
 at the present epoch. This result  is consistent with conclusions
 arrived
 at by Bertolami and Martins $[8]$, N.Banerjee and D.Pavon $[9]$,Sen and Seshadri
 $[10]$ that  $\omega$ should possess a low negative value for a satisfactory explanation of  structure formation,cosmic acceleration,
 coincidence problem,  and to avoid the problems of quintessence etc within the formalism of B-D theory. 
\newpage
\section{ DISCUSSIONS AND CONCLUSIONS}.

          In this work,we wish to emphasize that we have, for the first time derived the  explicit time dependence
          of the Brans-Dicke parameter  $\omega$ by solving gravitational 
 field    and wave equations of generalised B-D theory consistently, assuming power law
 behaviour for the scale factor
$a(t)$ and scalar field $\phi(t)$.Interestingly,we find two consistent
solutions of the  field and wave  equations.One solution  leads to
General Relativity with the implication  that  B-D theory is a more
general formulation than  GR. The  other solution, which is of greater
interest to us,leads to a time-dependent $\omega(t)$  whose time
dependence is governed by the  EOS parameter $\gamma$. Consequently,
$\omega$  exhibits different temporal behaviour in different epochs of
the evolving Universe characterised by its dominant matter/radiation
component. This,we believe,is an important result which can be used to study various characteristics of an evolving Universe within the generalised B-D formalism. In particular,for an accelerated expanding universe,the present
value of  $\omega$ comes out to be negative  with a minimum value of
$-\frac{1}{2}$. This result, once again,  nicely agrees with the conclusions of the earlier works $[8,9,10]$ carried within the
formalism of constant- $\omega$  B-D theory  that
$\omega$  needs to be negative  for a successful explanations of the
various observed characteristic of the evolving universe.\\ 

\vspace{0.7 in}

\centerline {\bf Acknowledgements}
\vspace{0.1 in}
The authors are grateful  to DST,Govt.of India for providing financial
support.The authors also thank Institute of Physics,Bhubaneswar,India,for
providing facility of  the computer centre.

\newpage


\end{document}